\documentstyle[epsfig,12pt]{article}

\begin{document}

\begin {center}
{\bf SYNCHRONISATION OF RESONANCES WITH THRESHOLDS}
\newline
\vskip 2mm
{D.\ V.\ Bugg\footnote{email: d.bugg@rl.ac.uk} \\
\
{\it Queen Mary, University of London, Mile End Rd., London E1\,4NS, UK}
\\[3mm]}
\end {center}

\begin{abstract}
The mechanism by which a resonance may be attracted to a sharp
threshold is described with several examples.
It involves a threshold cusp interfering constructively with either
or both (i) a resonance produced via confinement,
(ii) attractive $t$- and $u$-channel exchanges.
More generally, it is suggested that resonances are eigenstates
generated by mixing between confined states and long-range meson
and baryon exchanges.

\vskip 2mm

\end{abstract}

\section {Introduction}
Many examples are known where resonances lie close to sharp thresholds.
Table 1 lists several.
There is a simple explanation for this effect.
The rapid opening of a threshold causes a dispersive cusp at the
threshold, generating attraction there.
This attraction can capture a nearby resonance.

\begin{table}[htb]
\begin {center}
\begin{tabular}{ll}
\hline
Examples & Threshold \\
         & (MeV) \\\hline
$f_0(980)$ and $a_0(980) \to KK$ & 991 \\
$f_2(1565) \to \omega \omega$         & 1566 \\
$X(3872) \to \bar D(1865)D^*(2007)$   & 3872 \\
$Z(4430) \to D^*(2010)\bar D_1(2420)$ & 4430 \\
$Y(4660) \to \psi '(3686)f_0(980) $   & 4666  \\
$\Lambda_c(2940) \to D^*(2007)N   $   & 2945  \\
$P_{11}(1710),P_{13}(1720) \to \omega N$ & 1720 \\
$K_0(1430) \to K\eta ' $              & 1453   \\\hline
\end{tabular}
\caption{Likely examples of resonances locked to threshold cusps.}
\end {center}
\end{table}

A resonance denominator may be written
\begin {equation}
D(s) = M^2 - s - \sum _j \Pi _j ,
\end {equation}
where $M$ is the resonance mass and the last term is summed over
channels:
\begin {equation}
\rm {Im}\,\Pi_j = g_j^2\rho_j(s)FF^2(s);
\end {equation}
$g_j$ are coupling constants, $\rho_j$
are phase space factors and $FF(s)$ is a form factor in the amplitude
(which appears squared in the width of the resonance).
Amplitudes are analytic functions of $s$, so ${\rm Im}\, \Pi _j(s)$
is accompanied by a real part given by the dispersion integral
\begin {equation}
\rm {Re}\, \Pi _j(s) = \frac {1}{\pi} P
\int _{s_{thr}} ^\infty ds' \, \frac {\rm {Im}\, \Pi(s')} {s' - s},
\end {equation}
where $P$ denotes the principal value integral and $s_{thr}$ is the
value of $s$ at threshold.
Indeed, ${\rm {Im}\, \Pi (s)}$ itself originates from a contour integral
around the pole at $s' = s$.
The dispersive term $\rm {Re}\, \Pi (s)$ is equivalent to evaluating
the loop diagram for the open channel, e.g. $\pi \pi \to KK \to \pi
\pi$.

\begin{figure}[htb]
\begin{center} \vskip -14mm
\epsfig{file=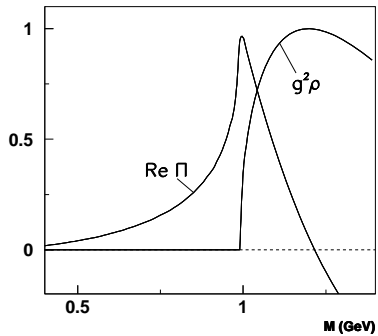,width=6.5cm}
\vskip -6mm
\caption {${\rm Re}\,  \Pi _{KK}(s)$ and $g^2_{KK}\rho _{KK}(s)$
for $f_0(980)$, normalised to 1 at the peak of $g^2_{KK}\rho _{KK}$.}
\end{center}
\end{figure}

Fig. 1 illustrates the $s$ dependence of $\rm {Im}\, \Pi (s)$ and
$\rm {Re}\, \Pi (s)$ for $f_0(980)$ at the $KK$ threshold.
For this channel, $\rho_{KK}=2k_{KK}/\sqrt{s} = \sqrt {1 - 4m^2_K/s}$.
The form factor is evaluated assuming a Gaussian source in $r$,
which gives (in amplitude)
\begin {equation}
FF = \exp \left[ -\frac {1}{6}\left(\frac
{kR}{0.197321}\right)^2\right],
\end {equation}
where $k$ is the
momentum of kaons in their centre of mass in GeV/c and $R$ is in fm.
For Fig. 1, $FF^2$ is taken to be $\exp (-3k^2)$, corresponding to a
radius $R = 0.84$ fm.
Results depend fairly weakly on this radius.
Note, however, that some form factor is required to make the dispersion
integral converge, since $\rho \to 1$ as $s \to \infty$.

The peak in $\rm {Re}\, \Pi (s)$ corresponds to an effective attraction
peaking at threshold.
If short-range attraction or meson exchanges generate a resonance
nearby, the cusp can lock the resonance close to threshold.

The effectiveness of the cusp as an attractor is studied in Ref.
\cite {Jphysg}
using $f_0(980)$ as an example, and taking $M$, $g^2_{\pi \pi}$ and
$g^2_{KK}$ from values found by BES for $J/\Psi \to \phi \pi ^+\pi ^-$
and $\phi K^+K^-$ \cite {phipp}.
If the $g^2$ are fixed at BES values and $M$ is varied, there is a pole
for all values of $M$ in the range 500 to 1100 MeV.
For $M$ between 800 and 1100 MeV, the pole is always within 50 MeV
of the $KK$ threshold, demonstrating that the
threshold can move the pole a long way from $M$.

In addition to the cusp effect, there is a further source of attraction
to the opening threshold.
The deuteron has a long-range wave function $\psi \propto
\exp (-\sqrt {M'B}r)$, where $M'$ is the $NN$ reduced mass and $B$ is
the binding energy.
For a resonance like $f_0(980)$, the wave function extends to
$\infty$ when $B \to 0$, i.e. at the threshold.
The total energy contains a kinetic energy term
$\nabla ^2 \psi /2M'$, and potential energy must overcome this
zero-point energy to make a resonance.
The curvature $\nabla ^2\psi$ of the wave function is a minimum when
the wave function extends to $\infty$.

T\" ornqvist \cite {Tornqvist} gives a formula for the $KK$
component in the $f_0(980)$ wave function, which can be written 
\begin {equation}
\psi = \frac {|q\bar q q \bar q> +\sum _i [(d/ds)\rm
{Re}\, \Pi_i(s)]^{1/2} |K\bar K>} {1 + \sum_i (d/ds)\rm {Re}\,
\Pi_i(s)}.
\end {equation}
Integrating over the line-shape of the
resonance, at least 60\% of $\psi ^2$ is $KK$; for
$a_0(980)$ the corresponding figure is 35\%.

\begin{figure}[htb]
\begin{center}
\vskip -10mm
\epsfig{file=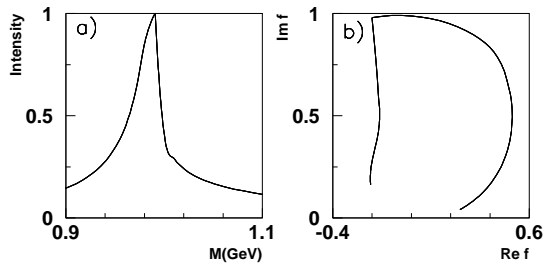,width=8.5cm}
\vskip -6mm
\caption {(a) the line-shape of $f_0(980)$, (b) its Argand diagram.}
\end{center}
\end{figure}

Fig. 2 shows the line-shape of $f_0(980)$ and its Argand diagram.
Note that the cusp effect at the $KK$ threshold is very strong, so
that the resonance is far from a Breit-Wigner of constant width.
The width 40-100 MeV for $f_0(980)$ quoted by the PDG refers
to the full-width at half-maximum.
For $f_0(980)$, $g^2_{\pi \pi}$ = 165 MeV and for $a_0(980)$,
$g^2_{\eta \pi} = 221$ MeV.
These are comparable with other resonances such as $f_2(1270)$.
In Fig. 2(b), the amplitude turns sharply through $90^\circ$ at the
$KK$ threshold as $ig^2\rho$ changes from real values below
threshold to imaginary values above.

Further algebra for the cusp is given in \cite {Jphysg}.
If one writes $\rm {Re}\, D(s) = M^2 - s + g^2m(s)$,
a power series expansion near threshold gives
\begin {eqnarray}
m(s)-m(s_{thr}) &=& -(2/\pi)\rho^2 + \ldots \qquad \qquad \qquad \qquad
{\rm above~threshold} \\
m(s)-m(s_{thr}) &=&-\sqrt {4m^2_K/s - 1} - (2/\pi) v^2 + \ldots
\qquad {\rm below~threshold}
\end {eqnarray}
where $v = 2|k|/\sqrt {s}$.
The first term in the latter expression is the usual Flatt\' e term for
the analytic continuation of $\rho _{KK}$ below threshold.
The next term in the expansion shows that the cusp is symmetrical to
first order in $|\rho|^2$ around the threshold.
The term $(-2/\pi)v^2$ may be rewritten $(2/\pi)(4m^2_K - s)/s$.
Note that this term resembles the term $M^2 - s$ in the expression for
$D(s)$.
Consequently $M$ and $g^2_{KK}$ are strongly correlated when fitting
experimental data; the BES group gives the correlation coefficients.
It is important to have data on the $KK$ channel to break the
correlation between parameters.

\section {Other examples}
\subsection {$X(3872)$}
Fig. 3(a) shows that the cusp due to the $\bar D D^*$ threshold
is too broad to account for the line-shape in the latest Belle data,
Fig. 3(b). A resonance or bound-state is required, as argued by
Braaten and collaborators \cite {Braaten} and Hanhart et al
\cite {Hanhart}.
The $\bar D D^*$ peak observed 3.5 MeV above the $\bar D D^*$
threshold may be fitted by folding the line-shape of $X(3872)$ with
$\bar D D^*$ phase space, as shown in Fig. 3(c).

\begin{figure}[htb]
\begin{center} \vskip -12mm
\epsfig{file=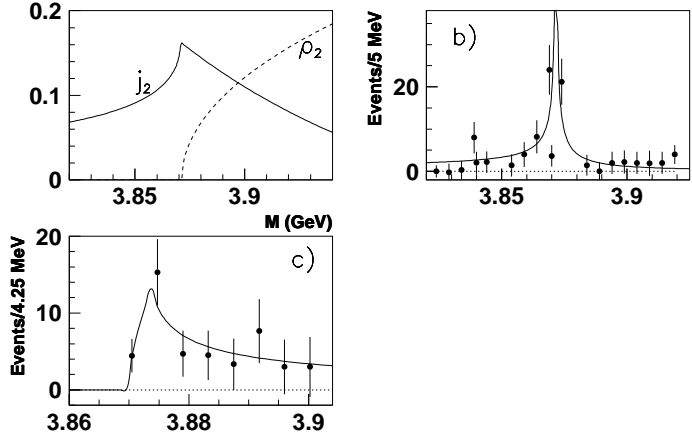,width=10cm}
\vskip -6mm
\caption {(a) $m_{\bar DD^*}$ and $\rho_{\bar DD^*}$ for
$X(3872)$; fits to Belle data [6] for (b) $X(3872) \to \rho J/\Psi$ and
(c) $X(3872) \to \bar D D^*$.}
\end{center}
\end{figure}

The $X(3872)$ could be a $c\bar c$ state which lies fortuitously
close to the $\bar D D^*$ threshold and has been captured by it.
There is however an $X(3940)$ reported by Belle in $\bar D D^*$,
though with only 24 events above background.
Confirmation is important. If all the high energy groups could
pool their data and determine the spin-parity, a value $J^P = 1^+$
for $X(3940)$ would require that $X(3872)$ is a molecule.
However, alternative possibilities are $J^P = 0^-$ and $2^-$,
i.e. $\bar D D^*$ P-wave states.

\subsection {$Z(4430)$}
A candidate for an exotic resonance is observed in $\psi '(3686)\pi
^\pm$ by Belle at the $D^*(2010)\bar D_1(2420)$ threshold, with a
width consistent within errors with that of $D_1(2420)$.
It can easily be fitted as a resonance.
However, it has many de-excitation modes, e.g. $\bar D D^*$ and
$\bar D^* D^*$ and can also be fitted within errors purely as a
threshold cusp \cite {Jphysg}.
The fit done this way is shown in Fig. 4(a) and the Argand diagram
in Fig. 4(b).
Telling the difference between a cusp and a resonance requires
interference with other components in the Dalitz plot.

\begin{figure}[htb]
\begin{center}
\vskip -12mm
\epsfig{file=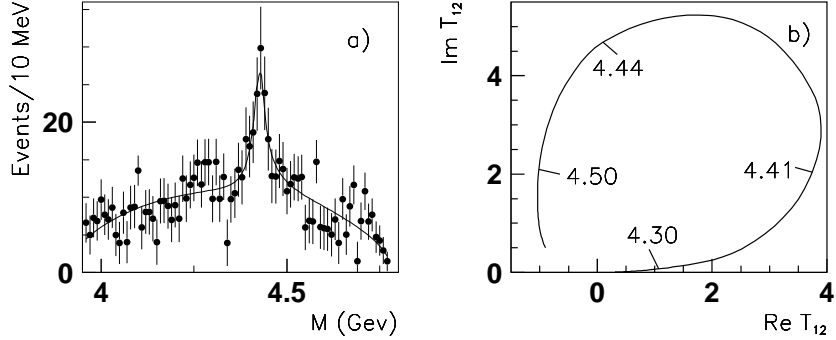,width=12cm}
\vskip -7mm
\caption {(a) Fit with a non-resonant cusp to Belle data for $Z(4430)$:
(b) Argand diagram.}
\end{center}
\end{figure}

\subsection {$f_2(1565)$}
The $f_2(1565)$ is definitely resonant.
Crystal Barrel data on $\bar pp \to 3\pi ^0$ trace out the phase as
a function of mass \cite {f01370}.
A sharp peak appears in $\pi \pi$ decays at 1565 MeV and also in VES
data on $\eta \pi ^+\pi ^-$ \cite {VES}, both due to the cusp at the
$\omega \omega $ threshold.
In the $\omega \omega$ channel, a reasonable form factor reproduces the
peak observed at 1640 MeV by both GAMS and BES \cite {PDG}.
The $f_2(1640)$ cited by the PDG is simply the $\omega \omega$ decay
mode of $f_2(1565)$ \cite {Baker}.
Incidentally, the $f_2(1430)$ of the PDG may also be explained by an
interference between $f_2(1270)$ and $f_2(1565)$ \cite {f01370}.

\subsection {$\eta (1405)$ and $\eta (1475)$}
These are probably two decay modes of a single resonance, $\eta
(1440)$.
The $\eta (1475)$ appears in $KK\pi$ data, mostly as $KK^*$, but
with a small contribution from $K(K\pi)_S$.
The $KK^*$ decay has $L=1$, hence a dependence $k^3$ on momentum in
this channel.
For the central mass of $K^*(890)$, the $KK^*$ threshold is at
1390 MeV and the $KK^*$ phase space increases rapidly with mass.
When one allows for $KK^*$ phase space, the 1475 MeV peak may be fitted
with a resonance at 1440 MeV.

The $\eta (1405)$ (or $\eta (1415)$ if one takes Crystal Barrel data
\cite {Nana}), appears in $\eta \pi \pi$.
This decay mode may be explained by $KK^*$ decay to $KK\pi$, followed
by $KK$ rescattering through $a_0(980)$ above the $KK$ threshold.
This signal is seen clearly in DM2, Mark III and BES I data on $J/\Psi
\to \gamma (KK\pi)$.
The BES I data are fitted simultaneously to $\eta \pi \pi$ data through
the $\eta \pi$ decay of $a_0(980)$ \cite {BES1}.
A full analysis requires inclusion of the strong dispersive effect due
to the opening of the $KK^*$ channel.

\subsection {$Y(4660)$}
Guo et al make a strong case for relating this peak to the
$f_0(980)\Psi '(3686)$ threshold \cite {Meissner}.
They include the dispersive threshold effect.

\subsection {Wider thresholds}
A full treatment of $a_1(1260) \to \rho \pi$ and $\sigma \pi$ requires
inclusion of the dispersive effect of opening channels.
Likewise, $a_2(1320)$ is affected by the opening of the $[\rho \pi
]_{L=2}$ channel.
In both cases, a Breit-Wigner resonance of constant width is a first
approximation.
The $f_0(1370)$ was originally fitted as a Breit-Wigner of constant
width, but in more detail its parametrisation (and particularly its
elasticity) is sensitive to the opening of the $\sigma \sigma$ and
$\rho \rho$ thresholds \cite {f01370}.
A similar analysis of $a_0(1450)$ is in progress.

The exotic $\pi _1(1600)$ has been convincingly confirmed at Meson08 by
new Compass data on the $\rho \pi$ channel. At lower mass, there are
claims for an additional $\pi _1(1405)$.
An analysis of these data including the opening of the $b_1(1235)\pi$
and $f_1(1285)\pi$ thresholds is needed to distinguish between a
resonance and threshold cusps.

\section {What are $\sigma$, $\kappa$, $a_0(980)$ and $f_0(980)$?}
Jaffe \cite {Jaffe} proposes to interpret these states as a nonet
due to SU(3) $3\otimes \bar 3$ states made up of $qq$ and $\bar q
\bar q$ pairs.
There is evidence for diquark interactions experimentally and in
Lattice Gauge calculations so there is probably some element of
such pairs in the wave function.
However, there are disagreements with observed decay branching
ratios \cite {decays}.
The ratio $\Gamma [f_0(980) \to KK]/\Gamma [a_0(980) \to KK]$ is
predicted to be 0.93, but is experimentally $2.15 \pm 0.4$ and
the ratio $\Gamma [f_0(980) \to \eta \eta]/\Gamma [f_0(980) \to \pi
\pi]$ is also at least a factor 3 less than predicted. The obvious
explanation is a substantial $KK$ component in the wave function. More
serious is that the ratio $\Gamma [\kappa \to K\pi]/ \Gamma [\sigma \to
\pi \pi ]$ is at least 6 standard deviations higher than predicted.
The  broad $\kappa$ is only just resonant and falls apart easily.

There are at least three reasons to believe that this nonet is driven
largely by meson exchanges. Firstly, Caprini et al. \cite {Caprini}
successfully predict $\pi \pi$ phase shifts and the $\sigma$ pole from
the Roy equations; these equations include crossed channels exactly and
therefore depend entirely on meson exchanges except for subtractions
for scattering lengths.
Secondly, Janssen et al. successfully predicted $f_0(980)$ and
$a_0(980)$ on the basis of meson exchanges \cite {Janssen}.
Thirdly, the model of Rupp and van Beveren successfully accounts for
the entire nonet with a single coupling constant and SU(3)
coefficients \cite {Joint}.
It finds all four states arise from the continuum, i.e. from
non-confined states.
An interesting detail  of the model is that the movement of
the poles can easily be traced as a function of the coupling
constant.
The $a_0(980)$ is {\it not} attracted to the $\eta \pi$ threshold
because of the Adler zero nearby; this Adler zero is built into the
model. Instead it settles close to the $KK$ threshold, because the
Adler zero in this channel is distant, at $s = m^2_K/2$.

I shall adopt the view that this nonet is largely driven by meson
exchanges.
We know that meson exchanges also account for features of $t$- and $u$-
channel peaks fitted to Regge exchanges throughout the intermediate
(few GeV) mass range.
My conclusion is therefore that meson exchanges combine with
short-range gluonic attraction to form all or most of the resonances
other than the $\sigma$ nonet.
The evidence that thresholds attract resonances is consistent with this
picture.
Oset, Oller and collaborators have found \cite {Oller}
they can predict a number of states purely from meson exchanges as
`dynamically driven' resonances, e.g. $\Lambda (1405)$ and $a_1(1260)$.
Likewise, Hamilton and Donnachie found in 1965 \cite {Donnachie} that
meson and baryon exchanges have the right signs to help generate
$P_{33}$, $D_{13}$, $D_{15}$ and $F_{15}$ baryons.

Let us carry this argument a step further.
Suppose contributions to the Hamiltonian are $H_{11}$ from confined
gluons and $H_{22}$ from meson and baryon exchanges.
It is almost inevitable that there will be mixing between these two
sources, and there is experimental evidence for such mixing for
$f_0(1370)$ \cite {f01370}.
Then the eigenvalue equation is
\begin {equation}
H \Psi  = \left(
\begin {array} {cc}
H_{11} & V \\
V & H_{22}
\end {array}
\right) \Psi,
\end{equation}
where $V$ describes mixing.
The Variational Principle ensures the minimum value of $E$ is the
eigenstate.
The stronger of $H_{11}$ and $H_{22}$ gets pulled down.
For $\sigma$, $\kappa$, $a_0(980)$ and $f_0(980)$, long-range forces
operate through mesonic S-waves and are stronger at low masses than
confinement, which leads to $\bar qq$ $P$-states near the mass of
$f_2(1270)$.
Otherwise, most non-$q\bar q$ and non-$qqq$ states are pushed
up and become too broad to observe.
There is a close analogy with molecular physics and the
formation of the covalent bond in chemistry. There is a cooperative
effect between confinement and $s$- and $t$-channel exchanges in
forming the ground-state.

Valcarce, Vijande and Barnea have made an interesting study
of mixing between diquark and tetraquark configurations
\cite {Vijande}, though they do not specifically take the cusp
effect into account.
They find that tetraquark configurations all lie higher than diquark
configurations if they ignore attraction between diquark pairs.
There must be some short-range repulsion between like
quarks due to the Pauli principle, as in the short-range repulsion in
nucleon-nucleon physics.
This repulsion plays a crucial role in preventing nuclear matter from
collapsing (and maybe for mesons and baryons too).
\section *{Acknowledgements}
I wish to thank Prof. G. Rupp and Prof. E. van Beveren for extensive
discussions.
Use of their model led to important insight into the way poles move
near a threshold.
I also wish to thank Dr. Y. Kalashnikova for illuminating
comments on the Flatt\' e formula.

\end {document}